\documentclass[twoside]{article}
\usepackage{fleqn,espcrc2}

\usepackage{graphicx}
\usepackage[figuresright]{rotating}

\newcommand{\AmS}{{\protect\the\textfont2
  A\kern-.1667em\lower.5ex\hbox{M}\kern-.125emS}}

\title{Gauge independence of Abelian and monopole dominance} 

\author{Fumiyoshi Shoji\address{Research Institute for Information
Science and Education, Hiroshima University,\\
 1-7-1, Kagamiyama,
Higashi-Hiroshima, 739-8521, Japan}
        \thanks{\texttt{shoji@riise.hiroshima-u.ac.jp}}}

\newcommand{\be}{\begin{equation}}
\newcommand{\ee}{\end{equation}}
\newcommand{\bea}{\begin{eqnarray}}
\newcommand{\eea}{\end{eqnarray}}

\begin{document}

\begin{abstract}
We formulate a stochastic gauge fixing method to study the gauge
 dependence of Abelian projection.
In this method, one can change the gauge from the maximally
Abelian one to no gauge fixing continuously.
We have found that the linear part of the heavy quark potential
 from Abelian contribution depends little on the gauge parameter.
Similar results have been obtained for the monopole contribution 
part.We also investigate the gauge dependence of the length of 
monopole loop, which is known to be important for the confinement, and
monopole density.
These results suggest that the picture that monopole plays an 
important role for the confinement of QCD dose not depend on choice of 
the gauge. 
\vspace{1pc}
\end{abstract}

\maketitle

\section{Introduction}

The investigation of quark confinement mechanism is an important 
issue in non--perturbative region of QCD.
As a way to approach the problem, Abelian projection has given us
many remarkable results. 
For example, Abelian and monopole dominances for string tension have 
been reported by G.S.Bali.et.al\cite{bali}, and H.Shiba and T.Suzuki 
have found the infrared 
effective action of QCD in terms of monopole by inverse Monte--Calro
method\cite{siba}. The infrared effective action can explain the 
monopole condensation by 
energy--entropy analysis, and reproduce the 
string tension analytically\cite{cher}.

These results suggest that monopoles which are given after 
Abelian projection have an important role for the confinement of the quarks,
and support the conjecture which has been
proposed by G.t'Hooft and S.Mandelstam\cite{thooft,mand}.

But as is well known, Abelian projection as a procedure is explicitly gauge
dependent. When $SU(2)$ gauge group is reduced into $U(1)$, 
there are infinite ways to fix gauge, and such a remarkable results have 
been reported only in Maximally Abelian (MA) gauge. 
If the conjecture is physically meaningful 
and to understand QCD vacuum from the point of view of the monopoles 
is correct, these results should be gauge independent.

Recently, Ogilvie\cite{ogilvie} has developed a character
expansion for Abelian and found that gauge fixing is unnecessary, i.e.,
Abelian projection yields string tensions
of the underlying non-Abelian theory even {\it without} gauge
fixing. Essentially the same mechanism was observed by Ambj{\o}rn and Greensite
for $Z_2$ center projection of $SU(2)$ link variables\cite{ambjorn,greensite}.
Furthermore, by introducing a gauge fixing function 
$S_{gf}=\lambda \sum \mathrm{Tr} U_\mu(x)\sigma_3
U_\mu(x)^\dagger\sigma_3$, 
Ogilvie has also shown that the Abelian dominance for
the string tension occurs for small $\lambda$.
Hence he conjectures that Abelian dominance is gauge independent and that 
gauge fixing results in producing fat links for 
Wilson loop and is computationally advantageous for the measurements.

\section{Formulation}

At this stage, it is important and significant to clarify 
1) gauge (in)dependence of monopole sector and 
2) intermediate region between with and without gauge fixing.

Here we employ stochastic quantization with gauge fixing term 
which has been proposed by D.Zwanziger\cite{zwanziger}.
This method allows us to choose gauge fixing condition from exact 
gauge fixing to no gauge fixing continuously.
Therefore to apply this method to MA gauge fixing yields us a powerfull tool
to research gauge dependence problem of Abelian and monopole dominance.

For MA gauge fixing, we construct Langevin equation for 
link variables with respect to 
fictious time by analogy with Ref.\cite{mizu}:

%This method is based on Langevin equation which describes 
%stochastic processes in terms of fictious time\cite{parisi}.
%A compact lattice version of this equation with gauge fixing
%was proposed in Ref.\cite{mizu}:

\bea
U_\mu(x,\tau+\delta\tau)&=&\omega(x,\tau)^{\dagger}
{\rm exp}(if^a_\mu \sigma_a)
U_\mu(x,\tau)\nonumber\\
&& \times \omega(x+\hat{\mu},\tau),\\
f_\mu^a&=&-\frac{\partial S}{\partial
A^a_\mu}\delta\tau+\eta^a_\mu(x,\tau)
\sqrt{\delta\tau},\nonumber\\
\omega(x,\tau)&=&{\rm exp}(i\frac{\beta}{2 N_c\alpha}
\Delta^a_{\mathrm lat}(x,\tau) 
\sigma_a \delta\tau).\nonumber
\eea

In MA gauge, 
\bea
\Delta_{\mathrm lat}(x,\tau)&=&i[\sigma_3,X(x,\tau)]\nonumber\\
&=&2(X_2(x,\tau)\sigma_1-X_1(x,\tau)\sigma_2),
\eea
where
\bea
X(x,\tau)&=&\sum_{\mu}(U_\mu(x,\tau)\sigma_3U_\mu(x,\tau)^\dagger
\nonumber\\
&&-U_\mu(x-\hat{\mu},\tau)^\dagger\sigma_3U_\mu(x-\hat{\mu},\tau)),\nonumber\\
&=&\sum_i X_i(x,\tau)\sigma_i,
\eea
here $\alpha = 0 (\infty)$ corresponds to exact MA gauge (no gauge fixing).

Then we calculate Wilson loops contributed from non--Abelian, Abelian,
monopole and photon by following way.
$SU(2)$ elements can be decomposed into diagonal and off-diagonal parts after
Abelian projection, 
\begin{eqnarray}
U_\mu(x)&=&c_\mu(x)u_\mu(x),
\end{eqnarray}
where $c_\mu(x)$ is the off-diagonal part and $u_\mu(x)$ is the diagonal one,
\[ u_\mu(x) = \left(
        \begin{array}{@{\,}cc@{\,}}
        \exp(i \theta_\mu(x) ) & 0 \\
        0                      & \exp(-i \theta_\mu(x) ) 
        \end{array}
\right) . \]
\noindent
The diagonal part can be regarded as link variable of the remaining $U(1)$.
One can construct monopole currents from 
field strength of $U(1)$ links\cite{degrand}:
\begin{eqnarray}
\theta_{\mu\nu}(x)&=&\theta_\mu(x)+\theta_\nu(x+\hat{\mu})
-\theta_\mu(x+\hat{\nu})-\theta_\nu(x)\nonumber\\
&=&\bar{\theta}_{\mu\nu}+2\pi n_{\mu\nu}~~~
 (-\pi\leq\bar{\theta}_{\mu\nu}<\pi) , \nonumber\\
k_\mu(x)&=&\frac{1}{4\pi}\epsilon_{\mu\nu\rho\sigma}  
\partial_\nu \bar{\theta}_{\rho\sigma}(x) ,
\end{eqnarray}
here, $k_\mu(x)$ is called ``monopole current'' which makes closed loop
on 4--dimensional lattice, and monopole density is
defined by

\begin{equation}
 \rho = \frac{1}{N}\sum_{x,\mu}|k_\mu(x)|,
\end{equation}
here $N$ is a normalization factor.

Wilson loops from Abelian, monopole and photon
contributions can be calculated as in Ref.\cite{siba}:
\bea
W^{{\rm Abelian}}&=&{\rm exp}(-\frac{i}{2}\sum_{x,\mu,\nu} M_{\mu\nu}(x)
\theta_{\mu\nu}(x)),\\
W^{{\rm monopole}}&=&{\rm exp}(2\pi i\sum_{x,x',\alpha,\beta,\rho,\sigma}
k_\beta(s)D(x-x')\nonumber\\
&&\times \frac{1}{2}\epsilon_{\alpha\beta\rho\sigma}
\partial_\alpha M_{\rho\sigma}(x')),\\
W^{{\rm photon}}&=&{\rm exp}(-i\sum_{x,x',\mu,\nu}
\partial_\mu^-\theta_{\mu\nu}(x)D(x-x')\nonumber\\
&&\times J_\nu(x')),\\
&&J_\nu(x)=\partial^-_\mu M_{\mu\nu}(x),\nonumber
\eea
where $\partial$ is a lattice forward derivative, $\partial^-$ is a backward
derivative and $D(x-x')$ is the lattice Coulomb propagator.
$J_\nu$ is the external source of electric charge and
$M_{\mu\nu}$ has values $\pm1$ on the surface inside of Wilson loop.

We calculate the heavy quark potentials from non-Abelian, 
Abelian, monopole
and photon contributions from these Wilson loops 
and extract the string tensions from each contributions.

As an improved action to reduce finite lattice spacing effects,
we adopt the Iwasaki action\cite{iwasaki}.
We also adopt Runge--Kutta algorithm\cite{runge} for improvement of 
finite step size effects and smearing technique for noise reduction.

\section{Numerical Results}
Numerical simulations were performed on $8^3\times12$, 
$16^3\times24$ and $24^3\times32$ lattices with
$\beta=0.995$, $\alpha=0.1,0.25,0.5,1.0$, and  $\delta\tau=0.001,0.005,0.01$.
Measurements were done every 100--1000
Langevin time steps 
after 5000--50000 thermalization Langevin time steps.
The numbers of Langevin time steps for the thermalization were
determined by monitoring the functional $R$ which is maximized in MA gauge 
and Wilson loops.

\begin{figure}[htb]
\vspace*{-0.8cm}
\begin{center}
\includegraphics[width=6cm]{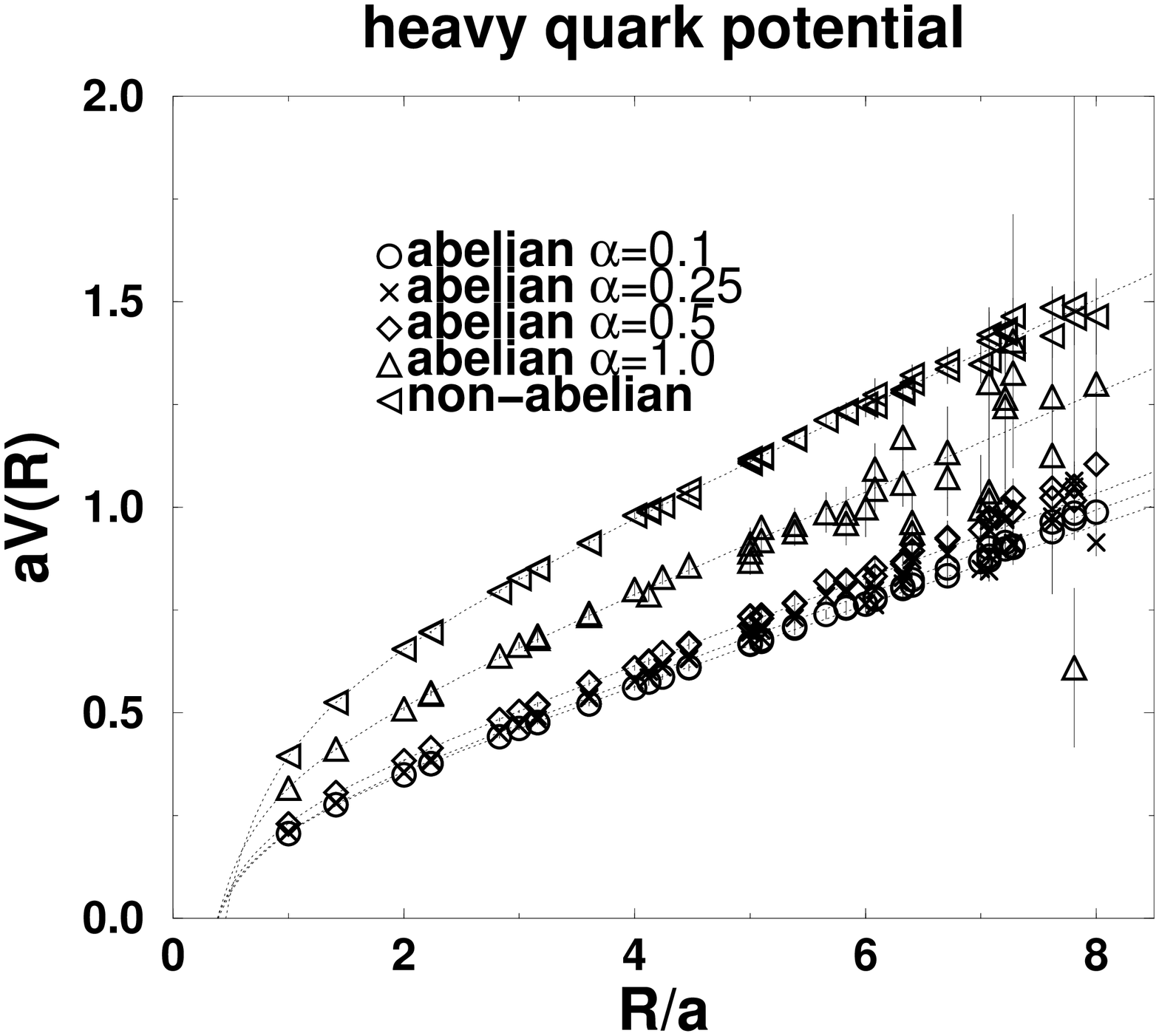}
\vspace*{-0.3cm}
\begin{minipage}{9.6cm}
\hspace{-0.8cm}
\includegraphics[width=5cm,height=4.cm]{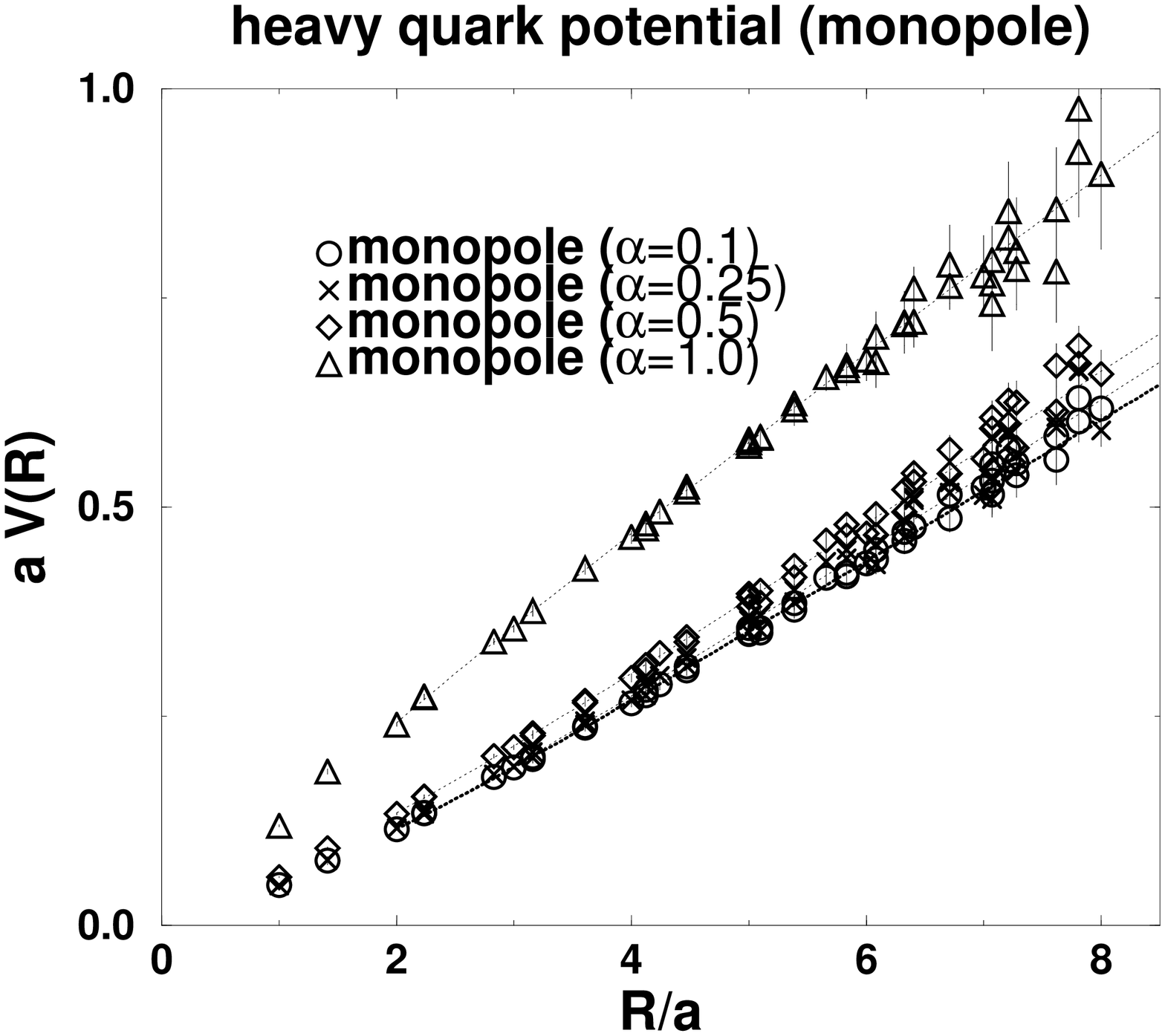}
\hspace{-0.8cm}
\includegraphics[width=5cm,height=4.cm]{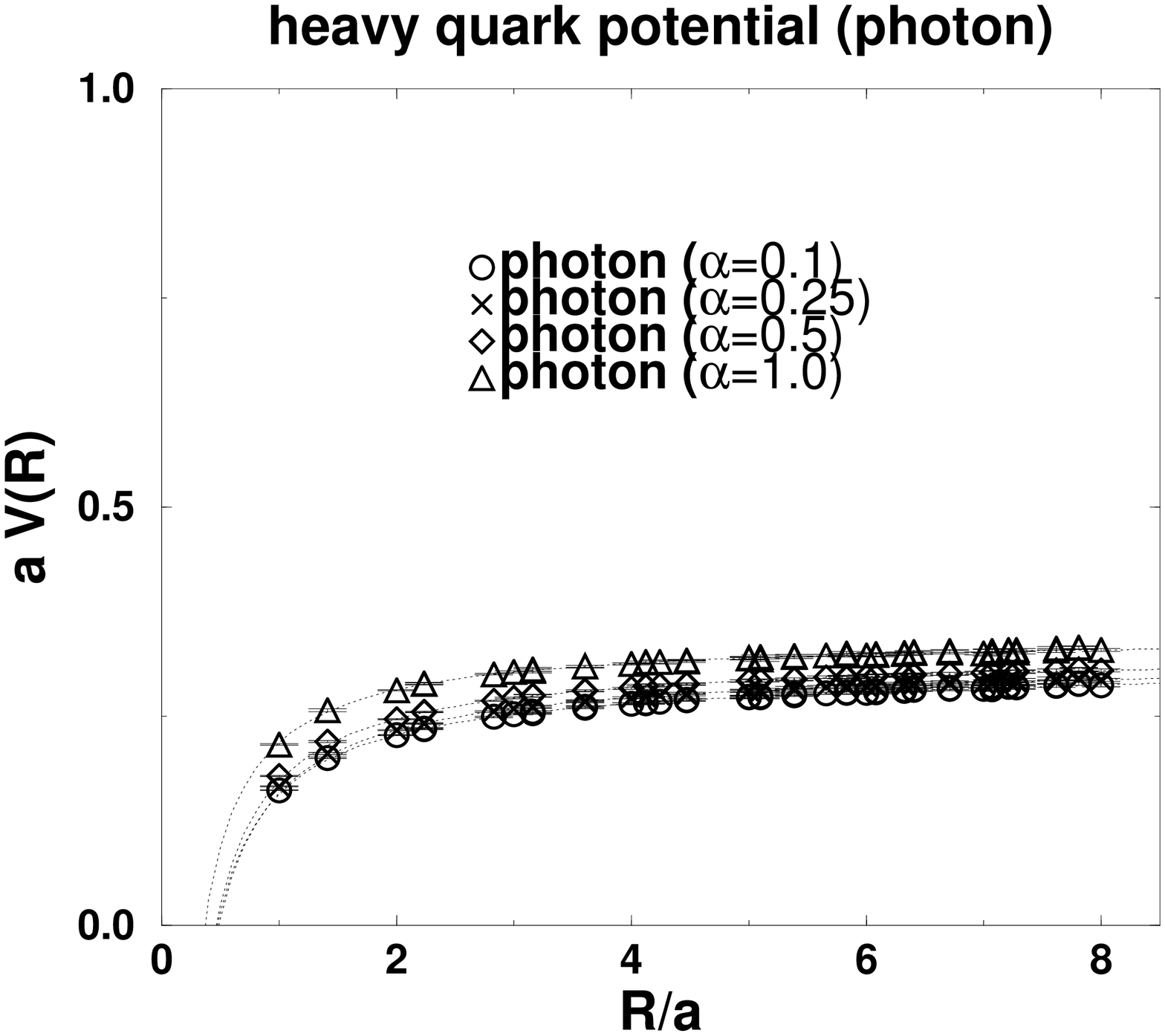}
\end{minipage}
\end{center}
\vspace{-0.8cm}
\caption{
Heavy quark potentials from non-Abelian, Abelian, monopole and photon 
contributions.
Lattice size is $16^3\times 24$, $\delta\tau=0.005$ and $\beta=0.995$.
}
\label{pot-abel}
\vspace{-0.5cm}
\end{figure}

%\vspace{-0.5cm}
In Fig.\ref{pot-abel} we show the heavy quark potentials from Abelian,
monopole and photon contributions 
for different $\alpha$'s
together with that of non-Abelian potential.
They can be well fitted by a linear and Coulomb terms, respectively.
We see that the linear parts of potentials are essentially 
same from $\alpha=0.1$ to $1.0$,
and all of them show the confinement linear potential behavior.
Therefore even when we deviate from the MA gauge fixing condition,
we can identify the monopole contribution of the heavy quark potential
showing the confinement behavior.
As $\alpha$ increases, statistical error becomes larger.
This result suggests that the gauge fixing is favorable for decreasing 
numerical errors as pointed by Ogilvie\cite{ogilvie}.

%\vspace{-1.cm}
\begin{figure}[htb]
\vspace*{-0.2cm}
\begin{center}
\includegraphics[width=7cm]{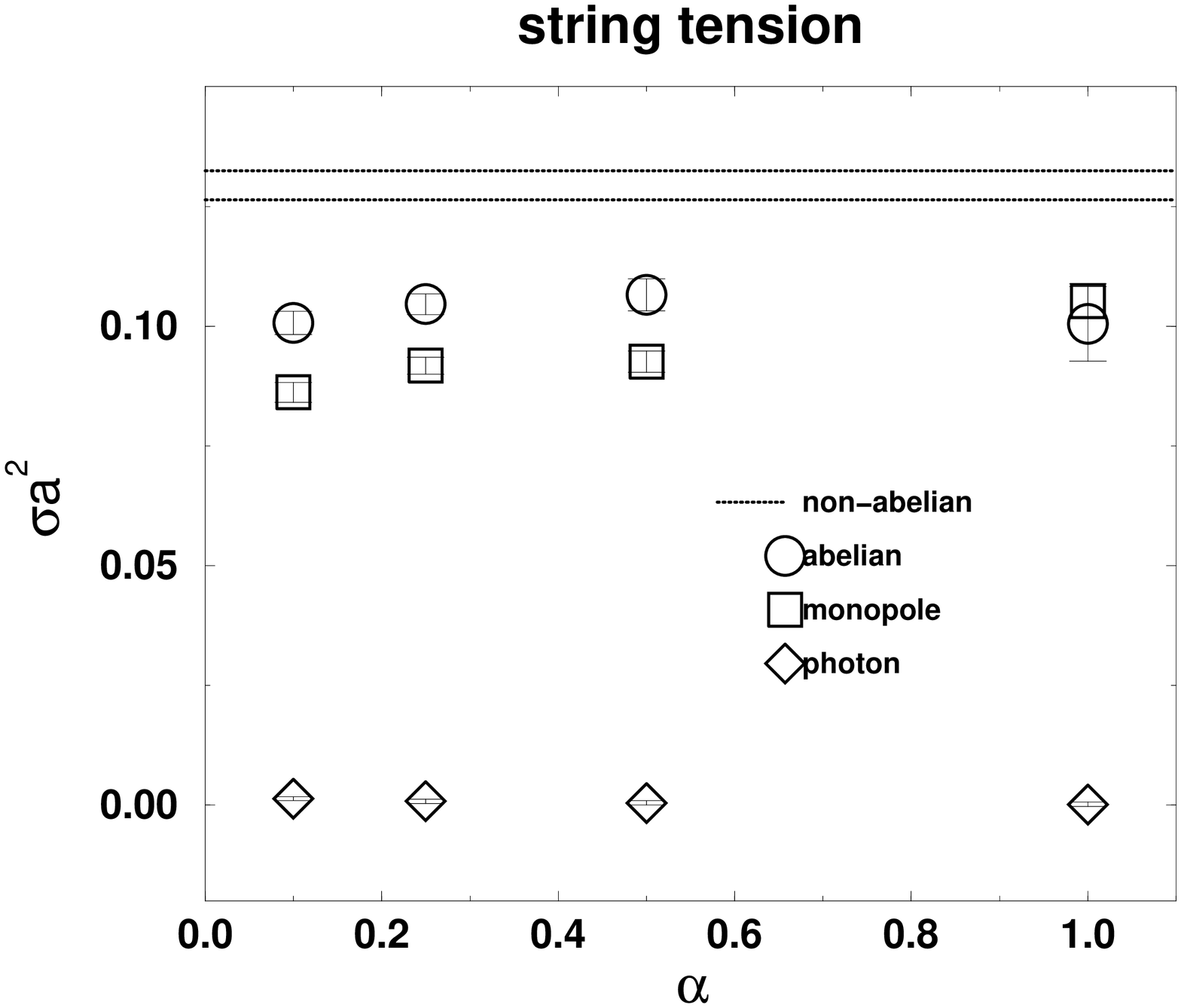}
\end{center}
\vspace{-1.5cm}
\caption{
String tensions from non-Abelian, Abelian, monopole and photon
 contributions.
Lattice size is $16^3\times24$, $\delta\tau=0.005$ and $\beta=0.995$.}
\label{string}
\vspace{-.6cm}
\end{figure}

%\vspace{-0.5cm}
In Fig.\ref{string} we plot the values of the string tensions from Abelian,
monopole and photon contributions as a function of the gauge parameter
$\alpha$.
They are obtained by fitting the data in the range $2.0\le R\le 7.0$.
We have taken into account only statistical errors.
The upper two lines stand for the range of the non-Abelian string
tension.
The Abelian and the monopole dominances are observed for all values of
$\alpha$.
On the other hands, the string tension from the photon part is consistent
with zero.
The string tensions from the Abelian parts are about 
80\% of the non--Abelian one. 
We expect that the difference of the percentage between our result and 
that of G.S.Bali et.al.\cite{bali} 
becomes smaller when we go to larger lattice size with proper noise
reduction technique.

%\vspace{-1.cm}
\begin{figure}[htb]
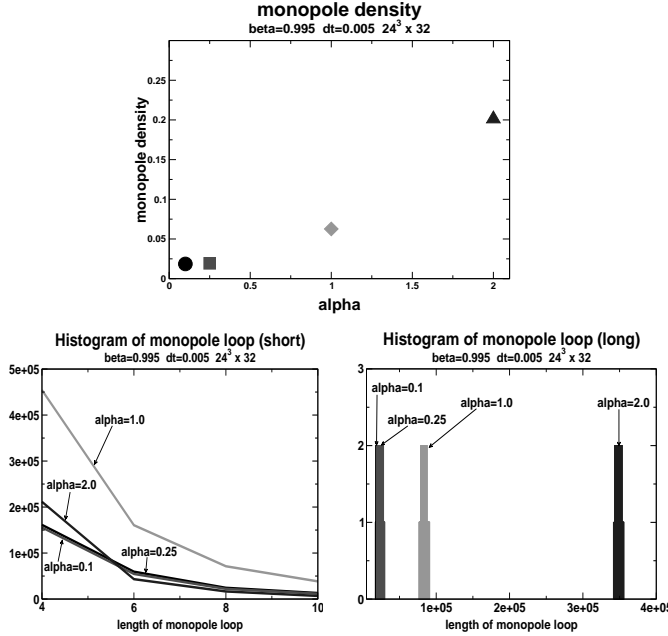

%\vspace*{-0.2cm}
\begin{center}
\includegraphics[width=5cm]{pics/dens_0.995_0.005.eps}
\hspace*{-.5cm}
\begin{minipage}{9.6cm}
\vspace{.2cm}
\includegraphics[width=4.2cm,height=4.cm]{pics/hist_s_0.995_0.005.eps}
\hspace*{0.2cm}
\includegraphics[width=4.2cm,height=4.cm]{pics/hist_l_0.995_0.005.eps}
\end{minipage}
\end{center}
\vspace{-1.2cm}
\caption{
Monopole densities versus several $\alpha$(top) and 
Histograms of short(left) and long(right) monopole loops.
Lattice size is $24^3\times32$, $\delta\tau=0.005$ and $\beta=0.995$.}
\label{dens}
\vspace*{-0.6cm}
\end{figure}

In Fig.\ref{dens}, we plot monopole
densities versus gauge parameter $\alpha$ and histograms of monopole loops.
When gauge fixing condition is far from MA gauge, 
monopole densities has larger value.
On the microscopic point of view, it has observed that short monopole
loops which make no cotribution to string tension increase drastically
as $\alpha$ increases. Therefore it become difficult to distinguish 
long monopole loop which is responsible for reproduction of string
tension from short one. 
Then the observables derived from monopole have large statistical error at
large $\alpha$ region.

\section{Concluding Remarks}

We have developed a stochastic gauge fixing method
which interpolates between the MA gauge and no gauge fixing.
The method is done together with the Iwasaki improved action.

We have studied the gauge dependence of heavy 
quark potentials derived from Abelian, monopole and photon contributions. 
For Abelian and monopole contribution, it is observed that the
confinement force is essentially independent of
the gauge parameter.
In the calculation of Abelian heavy quark potential, we have seen
that as gauge parameter $\alpha$ increases, 
the statistical error becomes larger.
This result suggests that the gauge fixing is favorable for increasing the
statistics as pointed by Ogilvie\cite{ogilvie}.
It is expected that as $\alpha$
increase, Abelian string tension would approach the non--Abelian one
\cite{ogilvie,greensite}.
%Therefore it is important to see behavior of the string tension
%as $\alpha$ becomes much larger than one.  
%But data are more noisy for large $\alpha$ and 
%we are planning to employ a noise reduction technique 
%such as integral method\cite{ppr}
%for obtaining statistically significant data.
But it is so difficult to measure heavy quark potentials in large
$\alpha$ region, because data are more noisy. 
In order to obtain statistically significant data even in such a region of
$\alpha$, more effective noise reduction technique such as integral
method\cite{ppr} will be need.

For monopole configurations, we have measured monopole density and 
histograms of monopole loops.
These results show when gauge condition is far from exact MA gauge
fixing, monopole configuration becomes very much complicated because of 
increase of short monopole loops and 
it is difficult to distinguish long monopole loop 
from short one.
Therefore in large $\alpha$ regions, heavy quark potential 
from monopole contribution has large statistical error.
If we apply block spin transformation in terms of monopole to such a
configuration, the clearer monoploe configuration would be obtaind.

\vspace{.3cm}
The author would like to thank to T.Suzuki at Kanazawa Univ. for 
many advices and suggestions and A.Nakamura at Hiroshima Univ. for 
fruitfull discussions.

\end{document}